\documentclass[preprintnumbers,amsmath,amssymb]{revtex4}
\pdfoutput=1
\usepackage{amsmath,amssymb,color,epsfig,latexsym,natbib,graphicx,dcolumn, bm} 
\usepackage{xifthen}
\usepackage{lpic}
\usepackage{caption}
\usepackage{floatrow}
\usepackage{subfigure}
\usepackage[normalem]{ulem}

\bibpunct{[}{]}{,}{n}{,}{,} 

\def\DEL#1{{\textcolor{green}{}}} 

\newcommand{\be}{\begin{equation}}
\newcommand{\ee}{\end{equation}}

\newcommand{\fig}[1]{Fig. \ref{#1}}
\newcommand{\figs}[2]{Fig. \ref{#1}-\ref{#2}}
\newcommand{\figp}[2]{Fig. \ref{#1}(#2)}
\newcommand{\eq}[1]{Eq. (\ref{#1})}

\begin{document}
\title{Evidence for Bolgiano-Obukhov scaling in rotating stratified turbulence using \\
high-resolution direct numerical simulations}

\author{D. Rosenberg$^1$, A. Pouquet$^{2}$,  R. Marino$^3$ and P.D. Mininni$^{4}$}
\affiliation{
$^1$National Center for Computational Sciences, Oak Ridge National Laboratory, P.O. Box 2008, Oak Ridge, TN 37831, USA.\\
$^2$Laboratory for Atmospheric and Space Physics, CU, Boulder, CO, 80309-256 USA.\\
$^3$NCAR, P.O. Box 3000, Boulder, Colorado 80307-3000, USA. \\
$^4$Departamento de F\'\i sica, Facultad de Ciencias Exactas y Naturales, 
Universidad de Buenos Aires, Ciudad Universitaria, 1428, Buenos Aires, Argentina.}
\begin{abstract}
We report results on rotating stratified turbulence in the absence of forcing, with large-scale isotropic initial conditions, 
using direct numerical simulations computed on  grids of up to $4096^3$ points. The Reynolds and Froude numbers are respectively 
equal to $Re=5.4\times 10^4$ and $Fr=0.0242$. The ratio of the Brunt-V\"ais\"al\"a to the inertial wave frequency, $N/f$,  is 
taken to be equal to 4.95,  a choice appropriate to model the dynamics of the southern abyssal ocean at mid latitudes. This gives a global 
buoyancy Reynolds number $R_B=ReFr^2=32$, a value sufficient for some isotropy to be recovered in the small scales beyond the 
Ozmidov scale, but still moderate enough that the intermediate scales where waves are prevalent are well resolved. We concentrate 
on the large-scale dynamics, for which we find a spectrum compatible with the Bolgiano-Obukhov scaling,
and confirm that the Froude number based on a typical vertical length scale is of order unity, with 
strong gradients in the vertical.  Two characteristic scales emerge from this computation, and are identified 
from sharp variations in the spectral distribution of either total energy or helicity. A spectral break is 
also observed at a scale at which the partition of energy between the kinetic and potential modes 
changes abruptly, and beyond which a Kolmogorov-like spectrum recovers. Large slanted 
layers are ubiquitous in the flow in the velocity and temperature fields, 
with local overturning events indicated by small Richardson numbers, and  
a small large-scale enhancement of energy directly attributable to the effect of rotation is 
also observed.
\end{abstract}
\maketitle

\section{Introduction} \label{S:intro}

Rotating stratified flows  are particularly important in the understanding of the dynamics of our planet 
and the Sun. Several of the key concepts needed in order to progress in predictions of the weather and 
in the global evolution of the climate depend crucially on a fundamental understanding of 
these flows. At different scales, different physical regimes become salient, and yet all scales 
interact. The nonlinear advection produces steepening, albeit slowly in the presence of strong waves. 
Thus, these fronts and turbulent eddies lead to enhanced dissipation and dispersion of particles and tracers, 
affecting the global energetic behavior of the atmosphere and climate systems, for example for atmospheric synoptic 
scales, and for oceanic currents, in the latter case modifying the meridional circulation. In the atmosphere,
such effects on energetics can in turn impair assessments of whether a given super-cell can
spawn a tornado, and they affect  both the evaluation of hurricane intensity and of climate 
variability. Rotating stratified turbulence (RST hereafter) thus plays a crucial role in the 
dynamics of the atmosphere and oceans, with nonlinear interactions--responsible for the complexity of 
turbulent flows--having to compete with the waves due to rotation and stratification. 

All of this takes place in the presence of a variety of other phenomena, including reactive chemical 
transport, biological or hydrological processes, as well as large-scale shear and bounday layers for example. 
One common approach is to tackle the problem in its entirety and construct a succession of models with 
increasing degrees of complexity. Conversely, one can take the simplest problem with what may be the 
most essential ingredients and examine the dynamics of such flows from a fundamental point of view, 
an approach taken in this paper. One of the inherent difficulties is the fact that such flows are 
represented, in the dry Boussinesq framework, by four independent dimensionless parameters, the Reynolds, 
Froude, Rossby and Prandtl numbers defined as:
\be
Re=\frac{U_0L_0}{\nu}, \ Fr=\frac{U_0}{L_0N}, \ Ro=\frac{U_0}{L_0f}, \ Pr=\frac{\nu}{\kappa} \ ,
\label{PARAM} \ee
where $U_0$ and $L_0$ are, respectively, a characteristic velocity and length scale, $\nu$ and $\kappa$ 
are the kinematic viscosity and scalar diffusivity (taken to be equal, $Pr=1$), $N$ is the  
Brunt-V\"ais\"al\"a frequency, and finally $f=2\Omega$ with $\Omega$ the rotation frequency. Other 
dimensionless parameters, combinations or variants of these basic ones, are commonly defined as well 
(see \S \ref{ss:param1}).

A number of studies have shown, at least in the absence of rotation, that the buoyancy Reynolds number 
$R_B=ReFr^2$ needs to be large enough for vigorous turbulence to develop in the small scales (see for 
example the review in \cite{ivey_08} and references therein). Indeed, at $R_B=1$, the Ozmidov scale 
\be
\ell_{OZ}=2\pi \sqrt{\varepsilon_V/N^3}, 
\label{OZM} \ee
at which isotropy recovers in a purely stratified flow, is comparable to the dissipation (or Kolmogorov) 
scale, $\ell_\eta=2\pi (\nu^3/\varepsilon_V)^{1/4}$, where $\varepsilon_V=|dE_V/dt|$ is the rate of dissipation 
of kinetic energy (note these length scales are written for a domain with dimensionless length of $2\pi$, 
such that $k=2\pi /\ell$ is the wavenumber). For $R_B>>1$, a Kolmogorov range, typical of 
isotropic and homogeneous turbulence, develops before dissipation can become effective. One can 
similarly define the Zeman scale, $\ell_{\Omega}=2\pi \sqrt{\varepsilon_V/f^3}$, for recovery of isotropy in 
a purely rotating flow, as shown in \cite{3072}.

According to the relative values of these parameters, several ranges can co-exist, with one effect 
overcoming others in each range (say, nonlinearities over wave motions or vice-versa). Thus, such 
flows support multi-scale interactions that have to be explicitly resolved. The interaction between 
oscillatory waves and steepening nonlinear interactions can also result, e.g., in the development 
of strong and localized vertical velocity fields \cite{rorai_14}. Different spectra are also observed in the purely 
stratified case; for example, a spectrum shallower than $k^{-1}$ is obtained in \cite{kimura_12}, whereas spectra steeper 
than $k^{-3}$ are observed in several other studies (see \cite{polzin} for a recent review of oceanic observations 
and analytical models). In both cases, non-local interactions between widely separated scales 
may well be dominant  \cite{lvov_12}. Thus, large scale separations have to be achieved in order 
to be able to unravel the different competing phenomena.

A high-resolution direct numerical simulation (DNS) of homogeneous isotropic turbulence on a grid 
of $4096^3$ points, with Taylor Reynolds numbers of up to 1200 was performed a decade ago 
\cite{kaneda, kaneda_rev} (for the case of passive tracers and Lagrangian particles, see 
\cite{sawford_11, sawford_13b}). For purely stratified flows, runs with a slightly smaller resolution were presented  
recently in \cite{almalkie_12}, with grids up to $4096^2 \times 2048$ points at the largest buoyancy 
Reynolds number, and $4096^2 \times 512$ for the more strongly stratified flow. In these simulations, 
energy cascades are found both in the vertical and the horizontal directions, with 1/3 of the 
dissipation coming from the former as in three-dimensional (3D) homogenous isotropic turbulence, and 
with a Kolmogorov spectrum in terms of the horizontal wavenumber at scales both larger and 
smaller than the Ozmidov scale. Other DNSs of purely stratified flows at linear resolutions of up to 
2048 points, at least in one direction, focus on the influence on the resulting dynamics and energy 
distribution among scales of resolving or not either the buoyancy scale characteristic of the thickness 
of the vertical layers
\be
L_B= 2\pi U_0/N \ , 
\label{LB} \ee
or the Ozmidov scale $\ell_{OZ}$ at which isotropy recovers \cite{waite2011,augier_12,bartello_13}. 
Part of the difficulty in determining spectral distribution among scales resides in the well-known 
fact \cite{cambon_89} that the dynamics is anisotropic, and thus the isotropic spectrum should 
be replaced by an axisymmetric two-dimensional spectrum, or by anisotropic correlation functions.
Similar characteristic length scales can be defined on the rotation rate $f$, and in fact, when both rotation and stratification are present, 
other scales can be defined 
(see equations (\ref{defBO}), (\ref{LBmod})).

It should be noted that numerical simulations are quite complementary to laboratory experiments. 
In the latter case, the Reynolds number can be quite high, reaching in some cases  geophysical 
values of $10^5$ or $10^6$, although Froude numbers often remain close to (but less than) unity 
\cite{ivey_91,barry_01}. This means that the buoyancy Reynolds numbers $R_B$ are high as well 
in these cases, although the stratification is not so strongly felt. By contrast, DNSs can only 
be performed at still modest values of Reynolds numbers (up to $\approx 10^4$, unless some 
parametrization scheme for the unresolved small-scales is used), but the Froude numbers can be 
taken as low as $10^{-2}$ or even $10^{-3}$ (for laboratory flows at small buoyancy Reynolds 
number, see the recent review in \cite{waite_14}).

\begin{figure*}
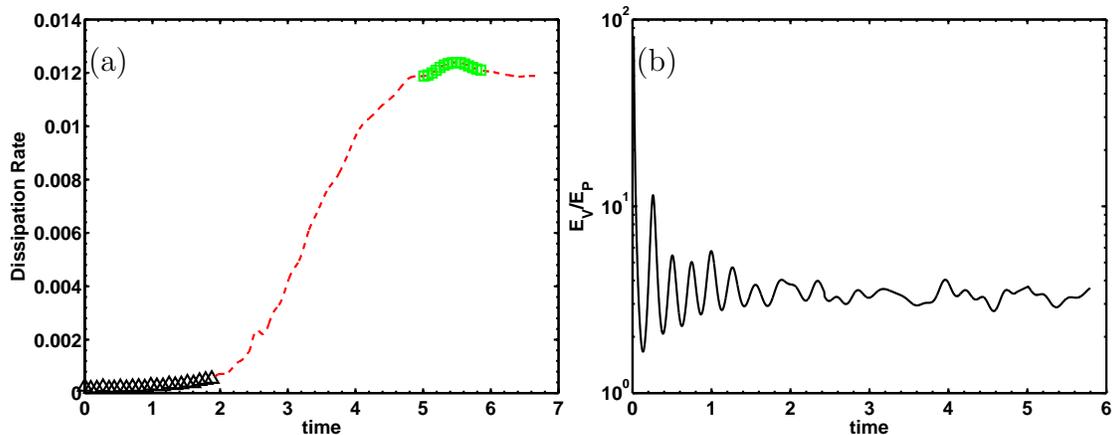
   
\begin{minipage}{0.45\textwidth}
\large
\begin{lpic}[]{fig1a(0.4,0.4)}
\lbl[l]{34,190;(a)}
\normalsize
\end{lpic}
\end{minipage}
\hspace{-1cm}
\begin{minipage}{0.45\textwidth}
\large
\begin{lpic}[]{fig1b(0.4,0.4)}
\lbl[l]{34,190;(b)}
\end{lpic}
\normalsize
\end{minipage}

\caption{Temporal variations of (a) kinetic energy dissipation rate  and (b) the ratio of kinetic to potential 
energy. In (a) is displayed with a dashed line (red) the run  using the $3072^3$ grid which 
evolved until $t=6.7$. The green squares represent the run performed on the grid of  $4096^3$ 
points, evolved for $5\le t \le 5.88$ (i.e., for a duration of $\approx 77$ gravity wave periods), and the black 
triangles indicate the early-time run on a grid of $1536^3$ points. All runs have the same physical parameters and time step.}
\label{compaenergy} \end{figure*}

While these results were obtained for purely stratified flows, the role of rotation on stratified 
turbulence has been investigated by a number of authors. Besides the energy, rotating 
stratified flows also conserve the pointwise  potential vorticity which can be defined as 
$P_V=f\partial_z \rho - N \omega_z + \omega \cdot \nabla \rho$, 
with $\rho$ the density (or temperature) fluctuations, and $\omega=\nabla \times {\bf u}$ the 
vorticity, ${\bf u}$ being the velocity. Because of the nonlinear term $\omega  \cdot \nabla \rho$ 
in the expression of $P_V$, its ${\cal L}_2$ norm is quartic and thus it is not conserved by each triadic 
interaction in a truncated ensemble of modes. The extent to which this is relevant to the dynamical 
evolution of the flow is not entirely known, but  several studies for shallow water \cite{warn_86} or the 
Boussinesq equations \cite{bartello_95, aluie_11, waite_13} assess the relative importance of the 
different contributions to $P_V$, with the general assumption that the high-order terms can be 
neglected when the waves are strong enough, i.e., at small Froude and/or Rossby numbers.  
In contrast, for the particular case of stable stratification, it was hypothesized in 
\cite{waite_13} that when $R_B$ is large enough the nonlinear term in $P_V$ affects the 
dynamics, becoming important at the same time as Kelvin-Helmoltz instabilities develop 
in the flow.

Since in many cases of geophysical interest, the ratio of the stratification to rotation frequencies 
$N/f$ is quite high (of the order of 100), most studies of RST consider the case of weak rotation. In 
reduced models relevant for geophysical flows, the geostrophic balance that results 
(between pressure gradients, Coriolis force and gravity) and the quasi-geostrophic (QG) regime, 
are central tenets of large-scale behavior and have been studied extensively over the years 
\cite{rhines_79,julien_12, klein_rev_10, vanneste_13}, including their breaking down through, 
for example, fronto-genesis \cite{molemaker_10a}.

\begin{figure*}
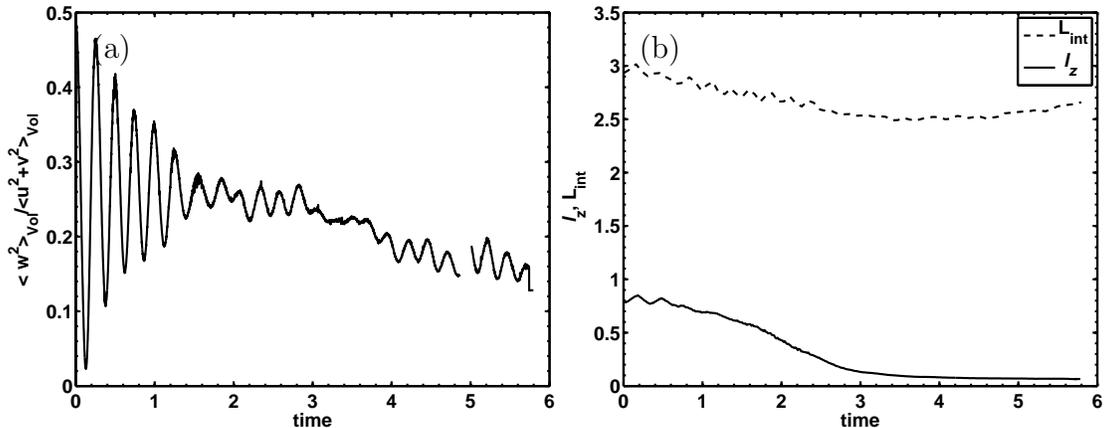

\begin{minipage}{0.45\textwidth}
\large
\begin{lpic}[]{fig2a(0.4,0.4)}
\lbl[l]{38,192;(a)}
\normalsize
\end{lpic}
\end{minipage}
\hspace{-1cm}
\begin{minipage}{0.45\textwidth}
\large
\begin{lpic}[]{fig2b(0.4,0.4)}
\lbl[l]{38,192;(b)}
\end{lpic}
\normalsize
\end{minipage}
\caption{Temporal evolution of (a) the ratio of the volume averaged vertical to horizontal 
kinetic energy, $\left<w^2\right> / \left<u^2+v^2\right>$, and (b) the 
vertical length scale $\ell_z$ defined in \eq{LZ}, which is characteristic of vertical shear layers. The integral 
scale $L_{int}$ is also provided in order to compare with $\ell_z$. 
} \label{fig_time} \end{figure*}

In the Boussinesq framework, a number of pioneering analyses of RST were performed in 
\cite{billant_01, lindborg2005, liechtenstein_05, liechtenstein_06b, waite_06, hanazaki_02, smith_02}.
The role played by the ratio $N/f$ in these flows is relevant although, in some ways, poorly understood. 
In \cite{billant_01} it was shown that, while stratification in the absence of rotation determines 
the vertical length scale $L_\parallel$ (basically, the buoyancy scale $L_B$ associated with the 
thickness of vertical layers, with a Froude number based on this vertical length scale of 
order unity) independently of the  horizontal scale, $L_\perp$, in RST this scale has a more
complex dependence on the buoyancy scale and on $N/f$, in which Rossby number is the chief
discriminating factor. However, specifically in the quasi-geostrophic
limit, it is found \cite{waite_06} that $L_\parallel \propto f L_\perp/N$, with the proportionality
indeed consisting of a function of Rossby number, as suggested in \cite{billant_01}. We use this
finding to help explain spectral features in our DNS.


In \cite{lindborg2005}, elongated boxes were considered to study the emergence of a direct energy 
cascade in RST with a Kolmogorov spectrum in the horizontal direction, and it was shown that such 
is the case provided the Rossby number is greater than a critical value of $\approx 0.1$. The case of large $N/f$ ($\gtrsim 45$) 
was also considered, and the runs were performed using 
hyper-viscosity. The aspect ratio of the computational domain seems to play an important role in 
these studies, and to influence the dynamics especially at unit Burger number $Bu=NL_\perp/fL_\parallel$. 
The linear regime of potential vorticity  at $Bu=1$ was analyzed in \cite{kurien_12} 
(see also \cite{remmel_10}), and it was found that vortical modes dominate over waves at large scales,
and that the parameter $\Gamma=f k_\parallel/(Nk_\perp)$ is relevant as a measure of the relative 
importance of terms in the linear part of the expression for potential vorticity: the two sources 
of dispersion become comparable when $f k_\parallel \sim N k_\perp$.  A more recent work on 
RST \cite{marino} deals with the emergence of helicity (vorticity-velocity correlations) in such 
flows, helicity being measured to be relatively strong in tornadoes and hurricanes \cite{moli1}, 
and also being an important ingredient in the origin of large-scale magnetic fields in astrophysics. 

Finally, besides DNS, rapid distortion theory for RST was considered in \cite{hanazaki_02} 
where it was shown that $N/f$ governs the final distribution of energy among the horizontal and vertical
kinetic energy components and  potential modes, as well as the normalized vertical flux
$\left<\rho w\right>$, where $w$ is the vertical velocity,
 together with the root mean square vertical vorticity, whereas 
stratification dominates the unsteadiness of these flows.

As already mentioned, $N/f$ is rather large in many applications. However, the case of RST 
with $N/f$ of order unity (or slightly larger) is also of interest for geophysical flows. One example is the abyssal 
southern ocean at mid latitude \cite{nikurashin_12},  which 
serves as a motivation for the present study and for which $N/f $ is estimated to be between roughly 5 and 10. Flows with $N/f$ ranging from $0.1$ to 10 were 
analyzed in \cite{liechtenstein_05, liechtenstein_06b}; all runs were spin-down with initial conditions 
at $k_0\approx 10$. These authors stressed the importance of computing for long times 
compared to both the inertial and stratified periods of the waves, because of what are 
called slow modes, i.e., modes with zero wave frequency, as already emphasized in \cite{smith_02} (see also \cite{herbert_14}. 
In \cite{smith_02}, it was also noted that energy builds up with time at small scales, the flow being strongly intermittent.
Previous studies in the regime of moderate $N/f$ also showed that the inverse cascade of energy to large 
scales is more efficient in the range $1/2\le N/f \le 2$ \cite{EPL}, when wave resonances disappear 
\cite{smith_02}. Moreover, when  forcing RST at small scales, it can be shown that there is a clear 
tendency towards a $-5/3$ spectrum for the inverse cascade, as the Reynolds number increases for fixed 
parameters, together with the existence of a dual energy cascade: to small scales with a positive 
and constant energy flux, and to large scales with again a constant but negative energy flux \cite{pouquet_13b}. 

Noticing the scarcity of high-resolution DNS for turbulence in the presence of both rotation and stratification    
to date, and considering the geophysical relevance of flows with moderate values of $N/f$, we thus now analyze 
results stemming from one such run with a numerical resolution using up to $4096^3$ grid points
at the peak of dissipation. In the next section are given the equations, the numerical procedure and
the overall parameters. Sections \S \ref{S:temp} and \S \ref{S:spec} provide, respectively, the temporal
and spectral dynamics of the flow, \S \ref{S:struct} describes the physical structures that develop, 
and finally, \S \ref{S:conclu} offers a brief discussion and our conclusions.

\begin{figure}
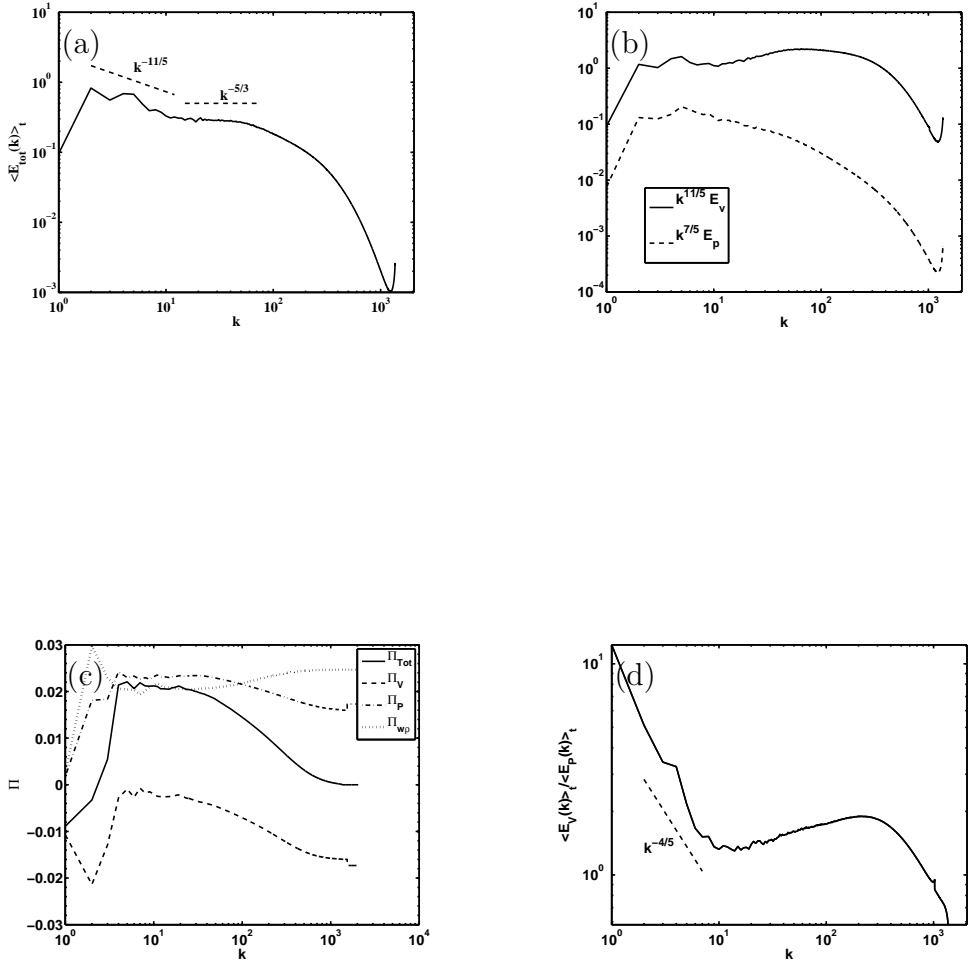

\begin{minipage}{0.45\textwidth}
\large
\begin{lpic}[]{fig3a(0.3,0.3)}
\lbl[l]{34,190;(a)}
\normalsize
\end{lpic}
\large
\begin{lpic}[]{fig3c(0.3,0.3)}
\lbl[l]{34,190;(c)}
\end{lpic}
\end{minipage}
\hspace{-1cm}
\begin{minipage}{0.45\textwidth}
\large
\begin{lpic}[]{fig3b(0.3,0.3)}
\lbl[l]{34,190;(b)}
\end{lpic}
\large
\begin{lpic}[]{fig3d(0.3,0.3)}
\lbl[l]{34,190;(d)}
\end{lpic}
\normalsize
\end{minipage}
\caption{
(a) High-resolution isotropic spectrum of the total energy, averaged over the time interval 
$t\in [5.3,5.7]$ corresponding to the peak in enstrophy, and compensated by a 
Kolmogorov 5/3 law. Note the break in the slope for $k\approx 12$. 
(b) 
Kinetic (solid line) and potential (dashed line) energy spectra compensated by $k^{11/5}$ and $k^{7/5}$, respectively, with the same temporal averaging.
(c)
Plot of total energy flux, and, separately, the kinetic and potential energy fluxes, as well as the 
buoyancy flux term obtained from \eq{buoyflux}. All fluxes are averated over the same time interval. 
Note the negative total flux at large scale, indicative of the effect of rotation.
(d)
Ratio of kinetic to potential energy spectra averaged over the same time interval; note again a transition around $k\approx 12$, and a scaling close to $k^{-4/5}$.
} \label{compaspec} \end{figure}

\section{Numerical set-up} \label{S:num}
\subsection{Equations } \label{SS:old}

The Boussinesq equations in the presence of solid body rotation, for a fluid with velocity ${\bf u}$, 
vertical velocity component $w$, and density (or temperature) fluctuations $\rho$, are: 
\begin{eqnarray}
\frac{\partial {\bf u}}{\partial t} + \mbox{\boldmath $\omega$} \times
  {\bf u} + 2 \mbox{\boldmath $\Omega$} \times {\bf u}  &=& -N \rho \hat e_z 
  - \nabla {\cal P} + \nu \nabla^2 {\bf u}  \ \ , \\
  \frac{\partial\rho}{\partial t} + {\bf u} \cdot \nabla \rho &=&  Nw + \kappa \nabla^2 \rho \ , 
\label{eq:momentum} \end{eqnarray}
together with $\nabla \cdot {\bf u} =0$ assuming incompressibility. ${\cal P}$ is the total pressure 
and $\hat e_z$ is the unit vector in the vertical direction which is in the direction of the imposed 
rotation and opposed to the imposed gravity; therefore, 
$\mbox{\boldmath $\Omega$} = \Omega \hat{z}$. The initial conditions for the velocity are 
centered on the large scales, with excited wavenumbers $k_0\in [2,3]$ and isotropic with random phases. 
In the absence of dissipation ($\nu=\eta=0$), the total energy $E_T=E_V+E_P$ is conserved, with 
$E_V=\frac{1}{2}\left< |{\bf u}|^2 \right>$ and $E_P=\frac{1}{2}\left< \rho^2 \right>$ 
respectively the kinetic and potential energies; the point-wise potential 
vorticity is also conserved. Lastly, $E_P=0$ initially. 

\begin{figure}
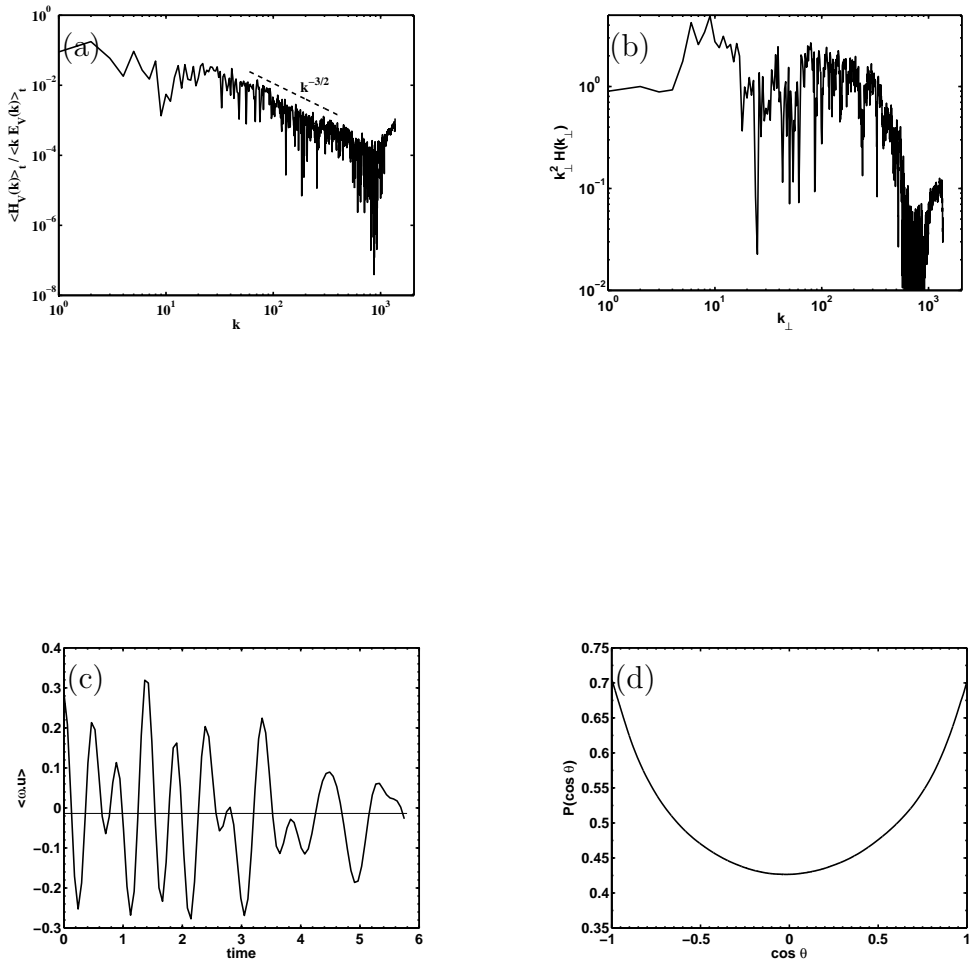

\begin{minipage}{0.45\textwidth}
\large
\begin{lpic}[]{fig4a(0.3,0.3)}
\lbl[l]{34,190;(a)}
\normalsize
\end{lpic}
\large
\begin{lpic}[]{fig4c(0.3,0.3)}
\lbl[l]{34,190;(c)}
\end{lpic}
\end{minipage}
\hspace{-1cm}
\begin{minipage}{0.45\textwidth}
\large
\begin{lpic}[]{fig4b(0.3,0.3)}
\lbl[l]{34,190;(b)}
\end{lpic}
\large
\begin{lpic}[]{fig4d(0.3,0.3)}
\lbl[l]{34,190;(d)}
\end{lpic}
\normalsize
\end{minipage}

\caption{
Helicity dynamics 
using the data from the $4096^3$ run.
(a) Relative helicity spectrum $|H_V(k)|/[kE_V(k)]$, which is seen as rather flat at large scale and 
decaying faster than $1/k$ at small scale.
(b) Perpendicular spectrum of the helicity compensated with $k_\perp^{2}$. 
Note the region of 
excess helicity for small wavenumbers followed, for $k>k_c$ with $k_c \approx 12$, by a drop in the amplitude of the 
compensated spectrum, and with fluctuations associated with rapid changes in sign of 
the helicity. For $k>300$, a sharp drop is observed.
(c) Temporal evolution of the volume integrated helicity.
(d) Probability distribution function of the relative helicity (cosine of the angle between velocity and vorticity) at the peak of dissipation, $t=5.54$. 
Alignment and anti-alignment of ${\bf u}$ and $\omega$ are equally likely, as in homogeneous isotopic turbulence. 
 } \label{helicity} \end{figure}

When linearizing the above equations in the absence of dissipation, one obtains inertia-gravity waves 
of frequency 
\be 
\omega_k= k^{-1} \sqrt{N^2k_\perp^2+f^2k_\parallel^2} \, ,
\label{dispersion}
\ee 
with $k=\sqrt{k_\perp^2+k_\parallel^2}$, $k_\perp=\sqrt{k_x^2+k_y^2}$, and $k_\parallel = k_z$, respectively  
the total, horizontal (or perpendicular), and vertical (or parallel) wavenumbers (see, e.g., 
\cite{bartello_95, sagaut_cambon_08}). Fourier spectra will be built-up from their axisymmetric 
counterparts defined from the two-point one-time velocity covariance $U({\bf k})$
(see, e.g., \cite{3072})
\begin{eqnarray}
e_V(|{\bf k}_{\perp}|,k_{\parallel})=
    \sum_{\substack{
          k_{\perp}\le |{\bf k}\times \hat {\bf z}| < k_{\perp}+1 \\
          k_{\parallel}\le k_z < k_{\parallel}+1}} U({\bf k}) 
  &  = \int U({\bf k}) |{\bf k}| \sin \theta d \phi = e(|{\bf k}|, \theta) = e(k, \theta) \ ;
\label{etheta} \end{eqnarray}
here $\phi$ is the longitude with respect to the $k_x$ axis and $\theta$ the co-latitude in Fourier 
space with respect to the vertical axis. The function $e_V({\bf k}_\perp,k_\parallel=0)$ may be regarded
as the spectrum of two-dimensional (2D) modes, having no vertical variation. Note that for an 
isotropic flow, at a given point ${\mathbf k}$ in wavenumber space, the ratio of the axisymmetric 
spectrum $e_V(|{\bf k}_{\perp}|,k_{\parallel})$ to the isotropic spectrum is $\sim 1/|{\bf k}|$
because the size of the volume element in the isotropic case contains an additional (integrating) 
factor of $|{\bf k}|$ compared to the axisymmetric case. Hence, if the axisymmetric spectrum 
behaves as $k_\perp^{-\alpha}$, then the corresponding isotropic scaling will be $k^{-\alpha+1}$.
The spectrum  $e_V(|{\bf k}_{\perp}|,k_{\parallel})$ can also be decomposed into the kinetic energy 
spectrum of the horizontal components (velocity components $u$ and $v$), and of the vertical kinetic 
energy (velocity component $w$): 
\be
 e_V(|{\bf k}_{\perp}|,k_{\parallel}) = e_\perp(|{\bf k}_{\perp}|,k_{\parallel}) +
e_\parallel(|{\bf k}_{\perp}|,k_{\parallel}) \, . 
\label{eee} \ee
In the following we will also consider the reduced perpendicular spectrum \cite{sen2}
\be
E_V(k_\perp) = \Sigma_{k_\parallel} e_V({\bf k}_\perp,k_\parallel)\, ,
\label{ekperp} 
\ee
the reduced parallel spectrum $E_V(k_\parallel)$ (which has a sum over $k_\perp$), and the 
spectrum representing the perpendicular energy of the strictly three-dimensional (3D)  modes:
\be
E_{3D}(k_\perp) = E_V(k_\perp) - e_V({\bf k}_\perp,k_\parallel=0) \, .
\label{ek3dperp} 
\ee
Similar definitions hold for the helicity and potential energy  spectra, 
$h_V({\bf k}_\perp,k_\parallel)$ and  $e_P({\bf k}_\perp,k_\parallel)$, their reduced forms,
$H_V({\bf k}_\perp)$ and $E_P({\bf k}_\perp)$, as well as their 3D expressions (i.e., the perpendicular 
spectra of the 3D modes), $H_{V, 3D}({\bf k}_\perp)$ and $E_{P, 3D}({\bf k}_\perp)$.
They will be analyzed in the following sections.

\begin{figure}
\begin{minipage}{0.5\textwidth}
\large
\begin{lpic}[]{fig5a(0.4,0.4)}
\lbl[l]{35,190;(a)}
\normalsize
\end{lpic}
\end{minipage}
\hspace{-1cm}
\begin{minipage}{0.5\textwidth}
\large
\begin{lpic}[]{fig5b(0.4,0.4)}
\lbl[l]{35,190;(b)}
\end{lpic}
\normalsize
\end{minipage}

\caption{
(a)   Ratio $E_{3D}(k_\perp)/e(k_\perp,k_\parallel=0)$ of spectral energy in 3D modes {\it versus} that in 2D modes. Note again the transitions for $k\approx 12$ and $k\approx 300$.
 (b)  Parallel spectrum of horizontal kinetic energy $e_\perp(k_\perp=0,k_\parallel)$ (solid line, see 
equation (\ref{eee})) and parallel spectrum of potential energy $e_P(k_\perp=0,k_\parallel)$ (dashed line), 
both  compensated by $k_\parallel^{-3}$. Power laws are indicated as references.  All spectra are averaged over the peak 
of enstrophy, $t \in [5.3,5.7]$. Note the small flat range 
at large scales in $e_\perp$, both ending with equipartition at $k\approx 12$.
} \label{other_spec} \end{figure}

\subsection{Specific numerical procedure} \label{SS:proc}

\begin{figure*}
\includegraphics[width=12.0cm]{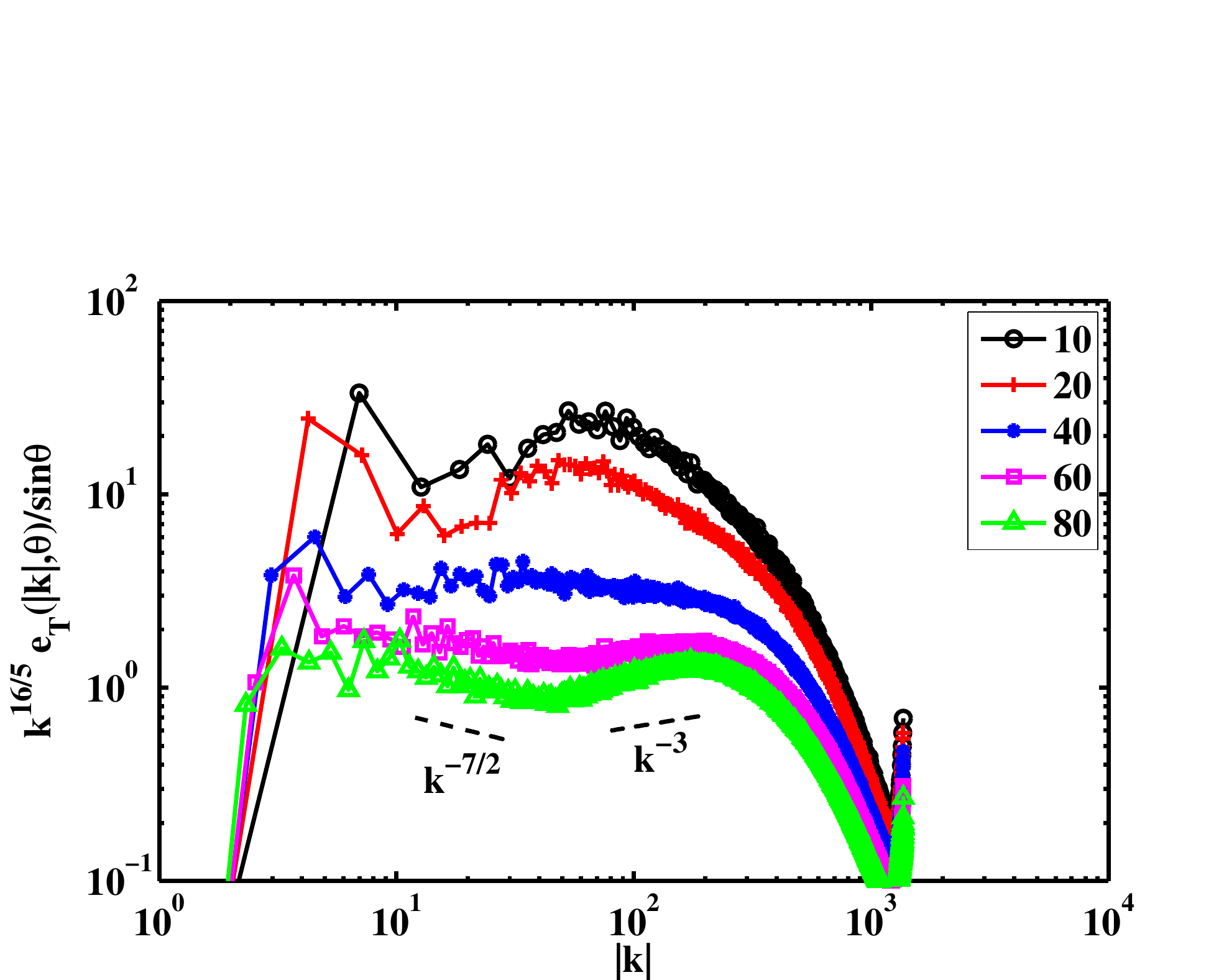}  
\caption{
Angular total energy spectra $e(|k|,\theta)$ (\eq{etheta}) for various co-latitude,
$\theta$, (in degrees) averaged over the peak of enstrophy, $t \in [5.3,5.7]$,
and compensated by $k(\theta)^{-16/5}$:
 $\theta=10^\circ$ (black circles),
 $\theta=20^\circ$ (red crosses),
 $\theta=40^\circ$ (blue asterisk),
 $\theta=60^\circ$ (magenta squares), and finally
 $\theta=80^\circ$ (green triangles).
The compensating slope corresponds to an (uncompensated)
isotropic (BO) scaling of $k^{-11/5}$. 
}
\label{ang_spec}
\end{figure*}

\begin{figure}
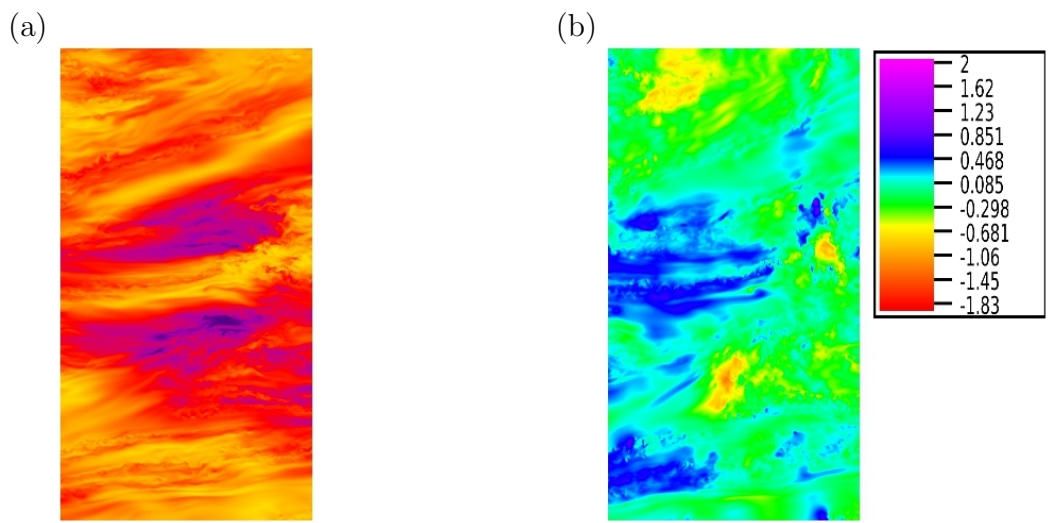

\hspace{-3cm}
\begin{minipage}{0.45\textwidth}
\large
\begin{lpic}[]{fig7a(0.4,0.4)}
\lbl[l]{80,200;(a)}
\normalsize
\end{lpic}
\end{minipage}
\hspace{-1cm}
\begin{minipage}{0.45\textwidth}
\large
\begin{lpic}[]{fig7b(0.4,0.4)}
\lbl[l]{80,200;(b)}
\end{lpic}
\normalsize
\end{minipage}
\vspace{-1cm}
\caption{
Perspective volume renderings of a thin y-z sub-volume of size
$0.4\times0.7$ times the compute box size at $t=5.54$ (close to 
the peak of enstrophy). 
The y-axis is directed horizontally, and the z-axis, vertically. 
Presented are (a) perpendicular and (b)
vertical velocity with identical color mapping. Note that the 
perpendicular velocity is dominant in magnitude. The slab thickness
in the x (depth) direction is $0.04$ times the box size. All renderings 
were made using the using the VAPOR visualization system  \cite{clyne07}.
}  \label{pvr1} \end{figure}

The code used in this paper is the Geophysical High Order Suite for Turbulence (GHOST), which 
is fully parallelized using a hybrid methodology \cite{hybrid2011}. It uses parallel multidimensional FFTs 
in a  pseudo-spectral method for 2D and 3D domains on regular structured grid, and can solve a variety 
of neutral-fluid partial differential equations, as well as several that include a magnetic field. 
Boundary conditions are periodic, and the time-integration is performed using a Runge-Kutta
algorithm up to 4th-order with double precision arithmetic. The code uses a ``slab'' (1D) domain 
decomposition among MPI tasks, and OpenMP threads  provide a second level of parallelization 
within each slab or MPI task. The code demonstrates good parallelization to more than $100,000$ 
compute cores.

In order to achieve a high resolution at peak of dissipation when gradients of variables are the 
strongest, we have implemented a ``bootstrapping'' procedure in which we start the simulation  at 
a lower resolution until the {\it dynamic range} of the energy spectrum decreases to some fiducial 
value. Here, by dynamic range we refer to the ratio of the energy at the peak of the spectrum, 
to the energy at the largest available wavenumber at a given resolution. When the lower threshold 
is reached, we increase the resolution and continue running until the dynamic range of 
the DNS at the new resolution decreases again to the fiducial value, repeating the process. 
Bootstrapping requires that a field at a reduced resolution be ``padded'' spectrally with zeros 
from its largest allowed wavenumber to the larger wavenumber allowed at the next (higher) resolution. 
This is handled in a processing step before the next highest resolution DNS is computed. This 
bootstrapping procedure was recently implemented, tested and used in the context of ideal 
magnetohydrodynamics \cite{brachet_13}.

We thus began with a $1536^3$ run up to $t=2$, then doubled the resolution on a grid of $3072^3$ 
grid points up to $t=5$, and then completed the run on the grid with $4096^3$ points. The maximum 
resolved wavenumber using a classical 2/3 de-aliasing rule is $k_{max}=N/3=1365$, with the length of 
the box corresponding to wavenumber $k_{min}=1$. The viscosity and scalar diffusivity were chosen
to be the same for these three successive runs, each run representing the evolution of the same 
physical problem at earlier times. The time step for each was chosen on the basis of the highest resolution 
considered in order to minimize time stepping errors at lower resolution.  The first bootstrapping 
was done during the inviscid phase before the small scale structures that can dissipate energy 
develop. 

The run on the intermediate grid of $3072^3$  points, was also pursued to later times
($t_{max}=6.7$); this enabled us to inspect the convergence of the overall statistics at the same
evolutionary times. Figure \ref{compaenergy}(a-b) displays the time evolution of the kinetic energy 
dissipation rate (proportional to the kinetic enstrophy $\left< |\omega|^2 \right>$), and the ratio of the kinetic to potential
energy, to illustrate the three distinct intervals with bootstraping and the overall 
evolution of the system.

\subsection{Other dimensionless parameters} \label{ss:param1}

As mentioned in the introduction, a variety of dimensionless combinations of relevant physical parameters 
can be defined for rotating stratified turbulence, beyond those written in \eq{PARAM}. One of the central 
limitations to a better understanding of such flows is the need to unravel what the key parameters are that 
govern the dynamics. Beyond the Reynolds, Froude, Rossby and Prandtl numbers, one also considers 
the ratio $N/f$, as well as the Froude number based on a characteristic vertical length scale, 
$$F_z=U_0/(\ell_ZN) \ .$$
Moreover, the combined effect of turbulent eddies and waves can be encompassed in the buoyancy and 
rotational Reynolds numbers, mentioned previously and respectively defined as 
\be
R_B=ReFr^2 , \ R_\Omega=ReRo^2 \ .
\label{RB} \ee
When $R_B\ge 1$ in a stratified flow, isotropy recovers beyond the so-called Ozmidov scale. Similarly, 
in a purely rotating flow, isotropy recovers beyond the Zeman scale for $R_\Omega\ge 1$ \cite{3072}.

The partition of energy between kinetic and potential modes can be measured  by their ratio, 
$E_V/E_P$, which is one possible definition of the Richardson number. Another definition is simply to 
measure the relative strength of the buoyancy to the inertial forces, or 
$$Ri=1/Fr^2 \ .$$
 However, in order to emphasize the role of the development of small scales in mixing, one can also  define a (local) Richardson number based on velocity gradients, $Ri_g$, as: 
 \begin{equation}
 Ri_g=  N(N-\partial_z \rho)/ (\partial_z u_\perp)^2  \ .  \label{eq:Ri} \end{equation}
 This definition suggests that a sufficiently large vertical gradient locally  leads to negative values of $Ri_g$, which is consistent with the intuitive picture of overturning when a denser parcel of fluid lies atop a less dense parcel.

\subsection{Run parameters and general characterization} \label{ss:param11}


\begin{figure}
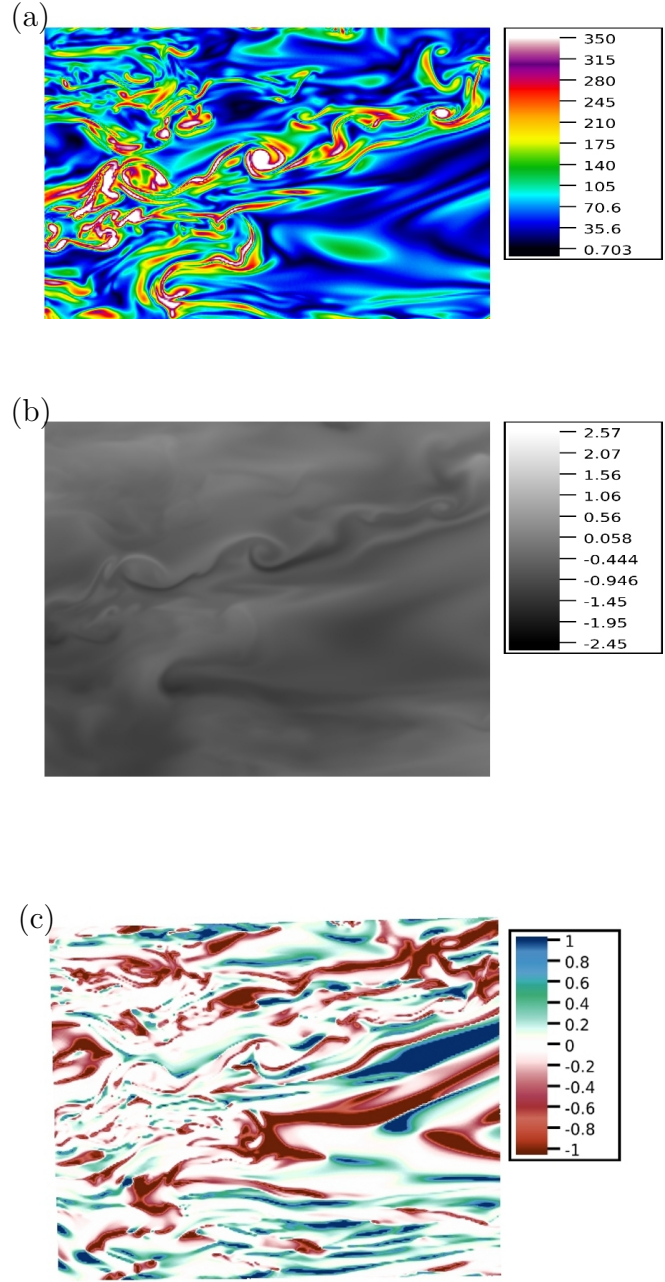

\large
\begin{lpic}[]{fig8a(0.27,0.27)}
\lbl[l]{70,240;(a)}
\normalsize
\end{lpic}
\vspace{-2.5cm}
\large
\begin{lpic}[]{fig8b(0.27,0.27)}
\lbl[l]{70,240;(b)}
\normalsize
\end{lpic}
\vspace{-1cm}
\large
\begin{lpic}[]{fig8c(0.27,0.27)}
\lbl[l]{59,240;(c)}
\end{lpic}
\normalsize
\caption{
Perspective volume renderings of a thin x-z sub-volume of 
size $0.12\times0.1$ times the compute box size at $t=5.54$ 
(close to the peak of dissipation); the  slab thickness in the y-direction is 0.01 times the box size.
The x-axis is directed horizontally, and the z-axis, vertically. 
Presented are 
(a) vorticity magnitude, 
(b) temperature fluctuations, and 
(c) local Richardson number $Ri_g$ defined in equation (\ref{eq:Ri}).
The color bar of vorticity illustrates the relatively intense vortices that are 
generated, and note the slanted Kelvin-Helmoltz layer. 
}
\label{pvr22} 
\end{figure}

We use $N/f=4.95$ with $N=13.2$ and $\Omega=f/2=1.33$ (thus, $f=2.66$). The viscosity is 
chosen to have the simulation well resolved: $\nu= 4\times 10^{-5}$. In dimensionless units, 
the resulting overall energetics of the flow lead to several scales that are of interest, and to a 
characterization of the flow in terms of the dimensionless parameters. Considered at the peak 
of enstrophy, the characteristic velocity is $U_0\approx 0.83$ and the integral length scale, 
computed from $L_{int} = 2\pi \int E_V(k)dk / \int k E_V(k)dk \approx 2.6$, very close as expected to the 
scale at which the energy spectrum initially peaks, namely  $L_0=2\pi/k_0\approx 2.5$. The dissipation rate 
of kinetic energy is taken from a computation of kinetic enstrophy at the peak of dissipation: 
$\varepsilon_V =\nu\left<|\omega|^2 \right> \approx 0.0124$ (see \figp{compaenergy}{b}). Note that in the isotropic case, 
$\varepsilon_V=\epsilon_{K41} = U_0^3/L_{int}  \approx 0.22$, but this relation does not hold in the highly anisotropic 
system we are investigating. Rather, we can take an estimate coming from weak turbulence, namely $\epsilon_{K41}*Fr\approx 0.005$, within a factor of two of the measured rate of energy dissipation. 
The Kolmogorov dissipation wavenumber is computed at the peak 
of dissipation to be 
$k_\eta \approx 660$. 
The Zeman and Ozmidov 
wavenumbers are therefore found to be, respectively, $k_{\Omega} \approx 39$ and 
$k_{OZ} \approx 431$. The buoyancy wavenumber is $k_B = 2\pi/L_B \approx 16$; the lack 
of scale separation between $k_{\Omega}$ and $k_B$ suggests that it will be difficult to 
distinguish as separate effects those due to rotation and those due to stratification. The 
Reynolds number is thus found to be $Re\approx5.4\times 10^4$, the Froude number 
$Fr \approx 0.0242$, and the Rossby number $Ro \approx 0.12$. Consequently, the 
buoyancy and rotational Reynolds numbers are  $R_B \approx 32$, and 
$R_\Omega \approx 775$. 
The Richardson number is determined to be 
$Ri \approx 1700$, so the flow is, indeed, found to be strongly stratified.

Finally, we can define a Taylor Reynolds number as $R_{\lambda}=U_0\lambda/\nu$, with 
$\lambda = 2\pi [\int E_V(k)dk / \int k^2 E_V(k)dk]^{1/2}$ the Taylor scale. In classical 
homogeneous isotropic turbulence (HIT) $R_{\lambda}$ measures the degree of development 
of small scales. At peak of dissipation, $\lambda\approx 0.31$, leading to a rather large 
$R_\lambda \approx 6400$, quite high compared to similar computations in HIT (e.g., 
$R_{\lambda}\approx 1200$ in a HIT run at similar grid resolution \cite{kaneda, kaneda_rev}). 
This is linked to the fact that, in the presence of strong waves, the transport of energy to small 
scales is hindered and not as efficient, and the energy spectrum becomes steeper at least at large scales, resulting in 
a larger Taylor scale for the same viscosity. It is worth noticing that in the atmosphere the 
Taylor Reynolds number is estimated to be $R_\lambda \approx 20000$, and it may be the 
case that realistic simulations of stratified and rotating atmospheric turbulence may be feasible 
in the near future as a result of this effect. Finally, note also that the value of $R_\lambda$ puts 
the present computation above the different thresholds in $R_{\lambda}$ identified in 
\cite{laval_03} for various instabilities to develop, as, e.g., for the growth of vertical shear 
and the growth of vertical energy.

 \begin{figure*}  \includegraphics[width=7.5cm]{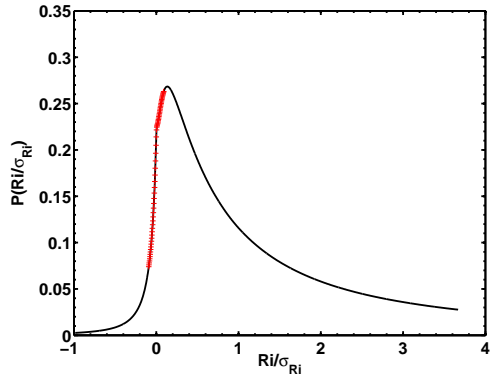} \caption{ 
Probability distribution function of the gradient Richardson number defined in \eq{eq:Ri}, at the 
latest time in the simulation. The (red)  crosses indicate where $|Ri_g| \le 0.25$, the classical 
criterion for  overturning instability \cite{miles_61,howard_61}.} \label{richardson} \end{figure*}

When the dimensionless numbers obtained in the simulation at peak of dissipation given above are now  
dimensionalized using the characteristic length and velocity of the abyssal southern ocean 
at mid latitudes, i.e. with $L_0=1000$ m (corresponding to the peak of energy input in the ocean from 
bathymetry \cite{scott_11}) and $U_0=0.024$ m s$^{-1}$, as measured for example in the Drake passage \cite{nikurashin_12}, we 
obtain kinematic viscosity and scalar diffusivity, respectively, of 
$\nu=\kappa=4.5 \times 10^{-4}$ m$^2$ s$^{-1}$, too large by roughly two orders of 
magnitude. The  corresponding overall effective energy dissipation rate would be 
$\epsilon \sim U_0^3/L_0 \approx 1.4 \times 10^{-8}$ m$^2$ s$^{-3}$; this latter value corresponds to the enhanced dissipation measured in the southern ocean \cite{naveira_04}. 
As a comparison, measurements in the atmosphere 
indicate $\epsilon \approx 10^{-6}$ m$^2$ s$^{-3}$  at intermediate 
altitude and at scales between 3 and $600$ km \cite{heas_12}. 
With  a rotation 
frequency of $\Omega = 10^{-4} \ s^{-1}$, our choice of parameters leads to a 
Brunt-V\"ais\"al\"a  frequency of $N \approx 10^{-3}$ s$^{-1}$, and $Fr\approx 0.024$, corresponding to the parameters of the run described above. Then, the  buoyancy scale is $150$ m, the Ozmidov scale is $4$ m, and the Kolmogorov 
dissipation scale is around $0.15$ m. This last value is  too large, because the 
viscosity is too large and the numerical resolution is still insufficient. Also, note that another lacking 
element in our simulation is the interaction with a larger-scale (mean) flow, say at the 
scale of several hundred kilometers, together with proper boundary conditions in the 
vertical.

\section{Overall temporal dynamics} \label{S:temp}

We now examine in more detail the overall temporal evolution of large-scale features. 
Figure \ref{compaenergy}(a-b) display, respectively,  the kinetic energy dissipation, 
$\nu \left< \omega^2 \right>$, and the ratio of kinetic to potential energy.
Easily identifiable initial oscillations due to the waves prevail at early times; these oscillations, 
stronger and thus more visible at large scale in the evolution of the energy, are due to inertia-gravity 
waves and their irregularity is linked with nonlinear coupling which, at that Reynolds number, is
sizable. However, the ratio of kinetic to potential energy remains relatively constant on average 
throughout the run after the initial phase, at a value close to 3. This initial phase is essential, 
since, even though our initial conditions have $E_P=0$ (and random phases for the velocity at large 
scale), the gravity waves provide a source of organized potential energy for the next temporal phase
when nonlinearities arise and constant-flux self-similar spectral scaling develops (see \S \ref{S:spec}).
The kinetic energy (not shown) 
starts to decay rather slowly as small scales have been formed. By the end of the run, the 
dissipation has reached a plateau and the flow is fully developed. When examining the temporal 
evolution of the energy and dissipation  for  the flows computed on $3072^3$ and  
$4096^3$ points, no differences are visible, indicative of a converged simulation and of a 
well-resolved flow. At the peak, $\varepsilon_V\approx 0.0124$, 
and the dissipation of potential energy is
$\varepsilon_P= \kappa \left< |\nabla \rho|^2 \right> \approx 0.0077$ (not shown).
 
In \fig{fig_time} are given the temporal evolution of the ratio of the ${\cal L}_2$ norms (volume 
averages) of the vertical to horizontal kinetic energy, as well as a characteristic vertical 
length scale defined as 
\be
\ell_z=[\left< u_\perp^2 \right> / \left< (\partial_z u_\perp)^2 \right>]^{1/2} \ .
\label{LZ} \ee
Note that $\ell_z$ can be viewed as a vertical Taylor scale, since it is based on vertical gradients of the velocity.
As expected, the horizontal energy dominates over the vertical at all times, by a factor 
close to 4, and increasingly so after the peak of enstrophy.  The vertical length-scale, of order 
unity to start with, undergoes a steady decrease and stabilizes as the peak of enstrophy is 
approached; it is one order of magnitude smaller at peak of dissipation when compared with 
its initial value. Considering now the vertical Froude number based on this vertical shearing 
length, $F_z=U_0/(N\ell_z)$, we find $F_z \approx 0.9 \lesssim 1$ at the latest time of the 
run. This value for $F_z$ is predicted for strongly stratified flows from the self-similarity 
analysis in \cite{billant_01}, if $\ell_z$ is taken to be the vertical scale of the dynamics, 
since, in this case, it is shown that $\ell_z \sim U_0/N$. One can contrast the anisotropy arising from rotation and stratification and say that the flow is fully turbulent but in an anisotropic manner \cite{marino_aniso}, although it still does feel the effect of rotation, as can be seen in \figp{compaspec}{c}, with a negative energy flux 
at large scale.

\section{Spectral behavior} \label{S:spec}
\subsection{Evidence for a large-scale Bolgiano-Obukhov scaling} \label{s:BO}

In \fig{compaspec} we show several isotropic spectra, which are all averaged averaged
around the peak of dissipation in the interval $t\in [5.3,5.7]$ (see \figp{compaenergy}{a}).
The total isotropic energy spectrum is compensated by a classical Kolmogorov $k^{-5/3}$ law. Such a 
law is compatible with the scaling of the spectrum observed at smaller scales, for 
$k_c \le k \le 100$ with $k_c\approx 12$; note that this value is
close to the buoyancy wavenumber $k_B\approx 16$ but may nevertheless differ from it (see below). 

At larger scales, a steeper spectrum is observed 
with a spectral slope close to $-11/5$, a value 
of 2.2 being computed from a least-squares fit on the interval 
$k\in[2,14]$). Note that spectra with a power-law index close to $-2$ were  found in 
\cite{kurien_14} for $N/f$ varying from 4 to 32, and observations in the ocean also 
indicate values that are similar and in fact closer to $2.5$ \cite{arbic_13}. 

One can invoke a dimensional argument to explain the large-scale spectral distribution, namely the  Bolgiano-Obukhov scaling 
(\cite{bolgiano_59, obukhov_59}; BO hereafter) derived for purely and stabley stratified turbulence. This scaling is obtained under the assumption that the source of energy at large scale is contained in the buoyancy, or in the potential modes, 
with nonlinear transfer rate $\varepsilon_P=|dE_P|/dt$, 
assumed constant, and with a negligible advection term in the momentum equation.
Since $\rho$  in the primitive equations written in \eq{eq:momentum} has the dimension of a velocity,
we have to re--introduce the physical dimension of the buoyancy flux in terms of length and time, i.e.,  $L^2 T^{-5}$; 
similarly one can use $\varepsilon_P N^2$ for the constant flux. This then 
leads to (see \cite{lohse_10} for a review):
\begin{equation}
E_V(k)\sim \varepsilon_P^{2/5} k^{-11/5} \ \ , \ \  E_P(k)\sim \varepsilon_P^{4/5} k^{-7/5} \ \ .
\label{eq:BO} \end{equation}
In the BO phenomenology, the scalar actively modifies the velocity field. 
Note that the Coriolis force does not contribute to the energy balance but only to an angular
redistribution of energy favoring negative flux to large scales, and thus does not perturb the 
dynamics leading to the BO scaling. The phenomenology derives from the idea that at large
scales, the nonlinear advection term is not strong enough
in the direct cascade to small scales, 
and the only available source of energy is therefore 
that coming from the scalar fluctuations. Requiring that the kinetic and potential energy spectra 
depend only on the dimensional buoyancy flux, $\varepsilon_P$, and wavenumber, $k$, leads to the 
above spectra. 

There are indications that the BO scaling has been observed  in stably stratified
in the atmosphere \cite{lovejoy_09}, as well as at the bottom boundary of convectively 
unstable cells, using temporal structure functions conditionally averaged on local values of the 
thermal dissipation rate \cite{ching_13}. A recent three-dimensional DNS analysis of Rayleigh-B\'enard 
convection shows such a scaling as well \cite{kumar_14}. BO scaling has been associated with 
a bi-dimensionalization of the flow due to stratification and the growth of the mixing layer
leading to a confined dynamics \cite{chertkov_03, boffetta_12}. In the case of the present computation, we note that
the quasi 2D large-scale dynamics is reinforced by the presence of rotation,
as observed in the kinetic energy flux which is negative, corresponding to  inverse transfer
(see below).

We show in \figp{compaspec}{b} the kinetic and potential energy spectra averaged 
over the time interval corresponding to the peak of enstrophy and compensated by the BO scaling.
This scaling seems to hold at large scales, up to $k\approx 12$ for the velocity, and on a shorter range for the temperature field.
In \figp{compaspec}{d} is shown the 
ratio of kinetic to potential energies, each averaged over time, and their ratio is consistent with a 
$k^{-4/5}$ law at large scale, 
as predicted by \eq{eq:BO} to within constants of order unity,
 whereas in the next regime, close to a Kolmogorov law, this ratio is close to equipartition in these units. 
\figp{compaspec}{c}  displays several fluxes.
The (forward) flux of total energy (solid line) is 
approximately constant, at a level of $\approx 0.022$ in these two identified ranges, indicative of a classical 
turbulent cascade. Note also that it becomes negative (reaching $\approx -0.0085$) at 
scales larger than the scale of the initial conditions; it can be expected, therefore, 
that, in the presence of forcing, a small inverse cascade may develop, as observed in 
\cite{aluie_11} and as it does when the forcing is placed at smaller scale (see e.g.,  
\cite{smith_96,EPL, pouquet_13b}). 

We also show in \figp{compaspec}{c} the energy flux decomposed into its  kinetic  
(dashed) and potential (dash-dotted) components, $\Pi_{V,P}$, as well as the buoyancy flux, $\Pi_{w\rho}$,
(dotted line), defined in wavenumber space as:
\be
\Pi_{w\rho}(k) = \sum_{k'=0}^{k'=k} \sum_{\, k'< |k''|<k'+1} \Re( \hat{w}(\mathbf{k}'') \hat{\rho}(\mathbf{k}'')^* )\,\, ,
\label{buoyflux}
\ee
where $\hat{w}(\mathbf{k})$ and $\hat{\rho}(\mathbf{k})$ are the Fourier coefficients for the 
vertical velocity and the scalar, respectively. 
The first two fluxes, $\Pi_{V,P}$, correspond to a scale--by--scale analysis of the 
two non-linear flux terms, $\rho {\bf u} \cdot \nabla \rho$ and ${\bf u} \cdot [{\bf u} \cdot \nabla] {\bf u}$, whereas the buoyancy flux concerns the energetic exchanges between the velocity and density fluctuations.
The sum of the kinetic enstrophy at its peak 
(see \figp{compaenergy}{a}) plus the kinetic energy flux, $\Pi_v(k=1)\approx -0.01$ is $\approx 0.0024$, which
is in excellent agreement with the nearly constant value of $\Pi_v$ in the region
$k\in[4,20]$ seen in this figure.  Furthermore, it can be seen that, as hypothesized in the BO 
phenomenology, the potential flux to small scales is  dominant, constant  and positive
for a wide range of scales. The kinetic flux has a strong peak at wavenumbers smaller than $k_0$. 
It is in fact negative throughout the wavenumber range around the peak of enstrophy;  this is likely 
due to the fact that the buoyancy flux acts as a source of energy for the velocity in 
a wide range of scales. 

We present the time average of $\Pi_{w\rho}$
in \fig{compaspec} (c;dotted curve), where it is seen that it is, in fact, 
comparable to the total energy flux, and can serve potentially as a kinetic energy source. 
We note that large temporal  fluctuations in the buoyancy flux are observed; they correspond 
to gravity waves directly affecting vertical motions.

Finally, we can evaluate the wavenumber, $K_{BO}$, at which the transition to 
a Kolmogorov spectrum $E_V(k)\sim \varepsilon_V^{2/3}k^{-5/3}$ is taking place,
in the framework of the BO scaling, by equating the two spectra at that scale. This leads immediately to 
\be
K_{BO}\sim \varepsilon_P^{3/4} \varepsilon_V^{-5/4} \ . 
\label{defBO} \ee
The value for $\varepsilon_P$ 
is taken to be that obtained in the large scales corresponding to the broad flat region in $\Pi_P$ observed in
\figp{compaspec}{c}; thus, $\varepsilon_P \approx 0.023$. For $\varepsilon_V$, we must be careful: this should be the
value that {\it would} be seen if we were able to resolve the Kolmogorov spectrum beyond the Ozmidov scale; 
however, this scale is barely resolved in this DNS. Hence,
we select the value of the kinetic energy flux at the largest wavenumber in the calculation to 
find $|\varepsilon_V|\approx0.015$. Using these values for the rates, we find that 
$K_{BO}\approx 11$, quite close to the observed value  
of $k_c\approx 12$.

The excellent agreement of the spectral scalings as well as the compatibility between the 
$K_{BO}$ computed with measured data and the observed $k_c$ offer compelling evidence of
BO scaling in this decaying strongly stratified, weakly rotating DNS. The problem remains, however, 
that there is little scale separation for $k<K_{BO}$ before a different dynamics dominates 
at larger wavenumbers.  A parametric study at high Reynolds number, achieved by varying the 
buoyancy force may help to determine the likelihood of such scaling laws in unbounded stratified 
turbulence; conditional averaging \cite{ching_13} may be effective for such a study.  

However, while shear is not imposed in our run, strong shear layers develop in the vertical in 
stably stratified flows, even in the presence of rotation (in which case they are slanted; see \figs{pvr1}{pvr22}). 
Shear is created locally and leads to strong instabilities (see \figs{pvr22}{richardson} below), so we must
consider its effect on spectral behavior.
A shear scaling leads to the following spectra:
 $$ E_V(k)\sim \epsilon_V^{1/3} S k^{-7/3} \ \ , \ \  E_P(k)\sim \epsilon_P  \epsilon_V^{-1/6} S^{-1/2} k^{-4/3} \ \ , $$
where $S$ is the shear rate (which can also be expressed in terms of a shear length scale) \cite{lohse_10}. 
In this case, the scalar is passive, and  the ratio of the two spectra varies as $k^{-1}$, so the spectral indices
are close to those that we find in our results. However, we have argued in part by considering $\Pi_{w\rho}$ (\eq{buoyflux})
and its magnitude relative to the total energy flux that the scalar field is not passive. Furthermore, the excellent agreement 
of the observed spectral indices and  the accord between the observed break in the spectra at $k_c$ and the computed
$K_{BO}$ seem to suggest that BO scaling is more likely; this may be a first instance of such a scaling in a DNS of 
strongly 
stably
stratified unbounded flows at relatively high Reynolds number (although, see \cite{kimura_96} and  \cite{kimura_12}).

\subsection{The lack of isotropy} \label{SS:aniso}

The transition in the spectral slope at $k_c\approx 12$ is not visible in the total energy 
flux; this was already noticed in \cite{3072} in the purely rotating case: even though 
characteristic time scales and nonlinear dynamics change with wavenumber, the flow 
of energy across scales is smooth. However, the wavenumber $k_c$
marks a clear transition in the character of the spectra, exhibiting also a sharp decrease of the ratio of kinetic 
to potential energy at large scales (see \figp{compaspec}{d}), followed by a quasi-equipartition 
between both energies for $k\ge k_c$ all the way to the dissipative scale (although with a slight variation 
with wavenumber). This change 
of behavior in the ratio of kinetic to potential energy at $k\approx k_c$ clearly indicates 
that wavenumbers $k\ge k_c$ corresponds to scales dominated by energetic exchanges 
between nonlinear eddies and wave modes eventually leading to the quasi-equipartition between 
kinetic and potential energy expected for strongly stratified flow \cite{billant_01}, 
while wavenumbers $k< k_c$ are sensitive to the effect of 
both buoyancy and rotation. Lastly, at 
the smallest scales of the flow dominated by dissipation processes, there is a broad 
decrease of kinetic energy compared to potential energy which is a likely a 
manifestation of overturning resolved in the small scales and leading to dissipative events and mixing (see also 
\fig{pvr22} below).

Moreover, in the presence of rotation and stratification, the flow loses its mirror symmetry. 
A measure of the departure from mirror symmetry can be obtained from the examination 
of the relative helicity spectrum, defined here in absolute value terms as:
\be
\sigma_V(k)=|H_V(k)|/[kE_V(k)] \ ,
\label{sigma} \ee
with $\sigma_V(k) \le 1 \ \forall k$ through a Schwarz inequality; $\sigma_V(k)$ is 
shown in \figp{helicity}{a}.  In HIT, $E(k)\sim k^{-e}$, and $H(k)\sim k^{-h}$ with 
$e=h=5/3$ so that $\sigma_V(k)\sim 1/k$ indicating a (slow) return to mirror 
symmetry in the small scales. In our case, the evolution is different: 
$\sigma_V(k)$ is rather flat for small wavenumbers, and decays as $\sim k^{-3/2}$ for 
wavenumbers larger than $k_c$. In the purely rotating case, it can be shown using 
dimensional arguments \cite{pouquet_10} that $e+h=4$, on the basis of a small-scale 
flux dominated by helicity which is an ideal invariant in that case (though not here). Assuming that 
the large-scale flow is dominated by rotation in a quasi-geostrophic regime, this 
leads to $e\approx 5/2$, close (but not identical) to the value found here for $k<k_c$, 
namely $e\approx 11/5$. It should be noted that this regime with $e=5/2$ corresponds to a fully 
helical flow ($\sigma_V(k)=1 \ \forall k$), a state which is known to be unstable 
\cite{podvigina_94}, and therefore an energy spectrum slightly shallower than
$k^{-5/2}$ should be expected instead. This energy spectrum (together with the flat 
spectrum of helicity) ends at a wavenumber $\approx k_c$, and one enters a rapid 
decrease of the helicity with wavenumber, slightly steeper than $1/k$, and with 
strong fluctuations likely corresponding to a rapid changes of sign in the helicity at 
various scales.

In \figp{helicity}{b} is presented the helicity spectrum $H(k_\perp)$ compensated with 
$k_\perp^{2}$. Note the region of 
excess helicity for small wavenumbers followed, for $k>k_c$ with $k_c \approx 12$, by a drop in the amplitude of the 
compensated spectrum, and with fluctuations associated with rapid changes in sign of 
the helicity. For $k>300$, a sharp drop is observed.
 Indeed, for wave numbers $k \lesssim k_c$ the compensated 
spectrum concentrates most of the helicity, which then decreases abruptly. This excess 
helicity at intermediate scales may derive from the alignment of the vortical structures 
produced by the rotation with vertical motions caused by buoyancy due to strong 
stratification, and may represent the physical mechanism for the generation of helicity 
proposed by \cite{Moffatt92,hide_76} and seen in direct numerical simulations in \cite{marino}.
In \figp{helicity}{c} we also show the temporal behavior of the
volume-averaged helicity. The flow starts with some residual positive
helicity (resulting from the random initial conditions), but after
$t\lesssim 4$ helicity fluctuates around zero.
 The lack of preference towards anti-alignment or alignment 
of velocity and vorticity can also be seen in \figp{helicity}{d}, which displays an average of PDFs of the cosine of the angle between velocity and vorticity. Note that  instantaneous PDFs (not shown) can display some 
slight excess  at $\pm 1$, corresponding to the fluctuations in the global helicity given in \figp{helicity}{c}.
 
In the presence of rotation and stratification, the flow also loses its isotropy. 
In \figp{other_spec}{a}, we  show the ratio of $E_{3D}(k_\perp)/e(k_\perp,k_\parallel=0)$, as 
defined in Eqs.~(\ref{etheta}) and (\ref{ek3dperp}). Both the numerator and denominator 
are averaged about the peak of dissipation on the time interval $t\in[5.3,5.7]$. This 
plot shows that at very large scales, there is roughly a constant and small amount of 
energy in the 3D modes compared with that in the 2D modes. Rotation
seems to play a role at these scales, mediating the distribution of kinetic energy 
between 2D and 3D modes, and accumulating more energy in 2D modes \cite{EPL}. As
larger  $k_\perp$ wavenumbers are considered, this distribution changes
rapidly until it reaches a local maximum around $k_B$. After a small decrease, the 
amount of energy in 3D modes far outpaces the distribution among 2D modes as 
expected in strongly stratified flows, as energy is transferred to large $k_\perp$ and 
potential modes are excited. In other words, the ratio
$E_{3D}(k_\perp)/e(k_\perp,k_\parallel=0)$ is consistent with a scenario in which the rotation, 
effective at large scales (presumably  for $k<k_\Omega$), controls the anisotropy, while 
at smaller scales as the system becomes dominated by stratification at the  buoyancy 
scale, $k_B\approx 16$, the energy is transferred towards modes with small $k_\perp$ 
but with $k_\parallel \ne 0$, resulting in most of the energy being in 3D modes.

According to \cite{billant_01}, under conditions of strong stratification ($Fr\to0$), the 
equations describing the flow become self similar. With rotation, self-similarity still 
holds, but the buoyancy scale \eq{LB} is suggested to take the modified form
\be
\tilde{L}_B = U_0 \mathcal{F}(Ro)/N \ = \ L_B \mathcal{F}(Ro) \, ,
\label{LBmod}
\ee
where $\mathcal{F}(Ro) \to 1$ when $Ro \to \infty$, and $\mathcal{F}(Ro) \to Ro^{-1}$ 
when $Ro \to 0$. 
In other words, under the effect of increasing rotation at fixed stratification, the  scale at which 
the effective Froude number in the vertical is of order unity increases as well, meaning that the large 
scales are more unstable. In the quasi-geostrophic (QG)  limit, for strong rotation and strong stratification, 
one can write that $N L_v/f= L_\perp$, a relationship that can be obtained simply, for example,  by equating in 
the dispersion relation the terms due to rotation and to stratification. This therefore defines a scale where 
rotation and stratification balance each other. Writing that $L_\perp$ is the integral scale $\approx 2.6$, 
we now find for the wavenumber where a change of behavior occurs between a rotation-dominated regime to a 
stratification dominated regime to be $\tilde{k}_B \approx 12$, 
a value that is in good agreement with $k_c$ 
as a break-point identified on several of the spectra presented here.    
To reconcile this with the evaluation of $k_{BO}$ given earlier, we could conjecture that the 
energetics of the flow at large scale is dominated by the buoyancy but the precise scale distribution 
of the energy is governed by the rotation as in the QG limit.

In \figp{other_spec}{b}, we also show plots of both $e_\perp(k_\perp=0,k_\parallel)$ and
of the spectrum of potential energy, both compensated by $k_\parallel^{-3}$, and shown 
at the peak of enstrophy. It has been predicted \cite{billant_01} that 
$e_\perp(k_\perp=0,k_\parallel) \propto k_\parallel^{-3}$, 
and similarly that the spectrum of the temperature fluctuations should also scale as 
$e_P \propto k_\parallel^{-3}$. 
The figure shows the existence of this prediction in the kinetic energy, but if such a range
exists in the potential energy, it is rather narrow. Both spectra 
seem to develop shallower power laws (other power laws are indicated in \figp{other_spec}{b} 
as references).  For $k>k_B$ the temperature and horizontal kinetic energy in these  
spectra are in approximate equipartition, which is expected for a self-similar range 
corresponding, in the primitive equations, to a balance between nonlinearity and wave 
dynamics. Note that a $k_\parallel^{-3}$ spectrum is often observed in the ocean, and is called the 
saturation spectrum; it is the regime in which, at least in the purely stratified case, 
intermittency of the vertical velocity is expected \cite{rorai_14}.

Lastly, in \fig{ang_spec} are shown the angular spectra for the {\it total} energy, 
(cf., \eq{etheta} for the kinetic energy) for several values 
of the co-latitude, $\theta$, i.e. the angle between the wave-vector ${\bf k}$ and the vertical.  
All spectra are averaged evenly around the peak of dissipation 
using ten temporal snapshots, and are compensated by $k_\perp^{-16/5}$, which is equivalent 
to compensating the isotropic spectra by $k^{-11/5}$ (see the discussion after \eq{etheta}). 
The angular spectra are computed by
interpolating the time-averaged 2D axisymmetric spectra along the line at a given co-latitude 
using a cubic interpolating polynomial. 
All scales are anisotropic, except close to the dissipative range; this is expected, since, 
in this simulation, $k_{OZ}\approx 431$ and  $k_{\eta}\approx 660$ (see \S \ref{ss:param11}).
Due to the dispersion relation, \eq{dispersion}, as $\theta\to 0$, inertial waves will dominate 
gravity waves, and as 
$\theta \to \pi/2$, the reverse will occur; the angular spectra reflect roughly a continuum in 
this behavior. The apparent tendency at small co-latitude for the spectrum to become very steep
at large scales suggests a quasi-two-dimensionalization due to strong rotational effects \cite{smith_02}.
At $\theta=20$, the steep range governed by strong rotation at the largest
scales gives way to a BO scaling at around $k\sim 10$, and the BO scaling range seems to spread to
larger scales as $\theta$ approaches intermediate values. But as the perpendicular direction is 
reached, multiple spectral ranges emerge after the BO scaling ends at the break-point $k=k_c\sim 12$ 
above.
In fact, a new characteristic scale seems to materialize at $k\sim 45$ for the largest co-latitudes
that may serve to separate distinct dynamical balances as illustrated by the reference slopes.

\section{Structures } \label{S:struct}

The salient physical structures that develop in this flow are relatively large, slanted layers, 
as can be seen in \fig{pvr1} displaying the horizontal and vertical velocity. The plots are 
perspective volume renderings of a thin y-z slab, and the dimensions of areas shown are 
$0.4\times0.7$ times the box size, comparable to the integral scale. The variation in the vertical 
direction is seen in these plots to be large, varying from filamentary-like thickness to structure at the
integral scale, 
which is comparable to the domain size. Additionally, in \fig{pvr22} are presented several 
renderings of a thin x-z  slab, zooming in on an area of $0.12 \times 0.1$ times the box 
size, comparable to the vertical Taylor scale. Note that $\ell_{OZ}$ is about $\frac{1}{3}$
of this slab size. These visualizations show scales at which overturning can occur and demonstrate the 
clear onset of Kelvin-Helmholtz instabilities due to shear layers. In both Figs. 
\ref{pvr1} and \ref{pvr22}, the thickness of the layers being visualized is $0.01$ in 
terms of the box size, roughly 1/6th of the Kolmogorov (dissipation) length.

The velocity is dominated by its perpendicular component, as already noted in \figp{fig_time}{a}.
As expected, the vorticity displays more small-scale variation (see \fig{pvr22}, left). A few 
large-scale vortices can be observed as well in the flow, but they are not visible in this sub-volume; they can be 
related to the role played by rotation, as already noted when examining the energy flux. 
The aspect ratio of the vortices has been found to depend  
on the global value of $N/f$ through, for example, the variation of correlation length scales 
\cite{lindborg2005, sukhatme_08}. It also depends on local values, as determined, for 
example, by the local rotation of the vortex \cite{aubert_12}.

In \fig{pvr22}, a clear vortex street appears at that time in  the vorticity (left),  the density (middle) and the gradient 
Richardson number (right) defined in \eq{eq:Ri},
showing that the flow can be locally unstable to overturning. 
Note the strong correlation between vorticity and temperature fluctuations, and the fact that 
the most unstable regions of the flow at this time are not strongly linked to the vortex street but that, 
in fact, other layers are being destabilized.
Note also the inter-mingling of stable and unstable structures at these scales. 
As mentioned earlier, the Richardson number based on velocity gradients 
(which can be defined in terms of $\ell_z$) can be considered as an overall index of the 
potential instability of the flow. 
A decrease in $\ell_z$ can thus be interpreted as leading to a more negative gradient Richarson
number, which is indicative of an evolution towards a flow more prone to overturning instability. 
Indeed, the probability distribution function of $Ri_g$ shown in \fig{richardson} indicates 
a strong probability of the flow meeting the classical criterion for overturning.
It was found in \cite{laval_03} that $Ri$ can become 
negative above $R_{\lambda} \approx 900$, with the change in sign coming from the 
change in sign of the vertical gradient of density. These results indicate that instabilities 
are triggered at various locations in the flow. In fact, actual bumps in the energy spectra 
have been observed in \cite{laval_03} at times of minima in the Richardson number for 
sufficiently high $R_{\lambda}$, that correspond to Kelvin-Helmholtz instabilities feeding 
directly the small scales.

\section{Conclusion} \label{S:conclu}

We have analyzed in this paper the results obtained from a high Reynolds number run of rotating stratified 
turbulence with $N/f=4.95$, characteristic of the abyssal southern ocean at mid latitudes. With a Froude 
number of $\approx 0.024$ and $Re\approx 5.5 \times10^4$, this run 
is 
not realistic in terms of 
Reynolds number
for geophysical fluid dynamics, and we have chosen to emphasize 
an examination of scales that are still dominated by the waves, with a barely resolved isotropic Kolmogorov 
range at small scales.  To unravel the role played by different phenomena, we examine the partition of several 
fields among scales.
We conclude that the largest scales (for $k < k_0$) are dominated by rotation, with a negative energy flux, and that 
for scales larger than a critical scale, $k_0< k < k_c$, the constant-flux range is one where the source of the 
energy is the potential energy stored in the large-scale gravity waves. We have presented evidence that
this energy source leads potentially to a Bolgiano-Obukhov scaling (\eq{eq:BO}). We have also demonstrated that 
this scaling is not necessarily inconsistent with the self--similarity argument of \cite{billant_01}.



The steep power-law observed at large scale is consistent with many oceanic observations, as analyzed for 
example in \cite{scott_05, arbic_13}.  The tendency for energy to pile-up in the large scales, even in the spin-down case, 
was already noted in \cite{metais_96}, where the inverse 
transfer was attributed to the geostrophic modes, whereas the wave modes undergo a direct energy cascade
(for a high-resolution forced case using hyper-viscosity, see  \cite{kitamura_06}).
At smaller scales, a Kolmogorov spectrum, in terms of horizontal wave numbers, obtains before isotropy is recovered, as already 
found in several studies of stratified flows. In addition to the conspicuous Kelvin-Helmoltz instabilities observed at
small scale, strong mixing at small scale is clearly favored as indicated both by an overall Froude number based on a 
vertical length scale of order unity, and by a PDF of the gradient Richardson number that shows directly the significant 
likelihood of overturning instability.

The regime with small Froude number and yet large buoyancy Reynolds number and moderate rotation, characteristic of many 
flows in geophysical fluid dynamics, 
remains a computational challenge, in particular when assessing highly non-local interactions between large scales fed by the inverse 
cascade of energy in the presence of rotation, even if weak, and small scales fed by the direct cascade of energy. Non-local 
interactions have been identified in such flows, for example in purely rotating flows \cite{alexrot}, in the context of the 
zig-zag instability \cite{deloncle_08}, and in rotating stratified turbulence \cite{aluie_11}. This clearly points out to the need 
of resolving the large-scale as well as the small-scale dynamics. In this regard, fundamental and idealized studies such as the one 
presented in this paper will remain valuable for some time to come, if only because they might lead to improved anisotropic and 
multi-scale parametrizations of such flows.

Many issues remain unexplored and one should  analyze in detail for example 
the distribution of energy among the normal modes of the flow (see e.g., \cite{bartello_95, sukhatme_08}), 
the small-scale behavior of the flow, and the role that helical coherent structures can 
play  in mixing, transport and intermittency in RST flows.
Indeed, helicity, or velocity-vorticity correlations is an ideal ($\nu=0$) invariant of the homogeneous isotropic case (as well 
as in the presence of solid body rotation), but when stratification is added, it can be created--as evidenced here--by quasi-geostrophic 
large-scale flows as a consequence of thermal winds \cite{hide_76, marino}. It is known that, for HIT in the presence of helical coherent 
structures, mixing is  modified. There are already sub-grid scale models of turbulence showing that, when taking helicity into 
account, the modeling capability is enhanced in a measurable fashion \cite{yokoi_93, baerenzung_11}, 
and thus the present study at high resolution may provide a useful database for testing a variety of parametrization schemes.

\begin{acknowledgments}
This work was supported by CMG/NSF grant  1025183,
and used resources of the Oak Ridge Leadership Computing 
Facility at the Oak Ridge National Laboratory, which is supported by 
the Office of Science of the U.S. Department of Energy under Contract 
No. DE-AC05-00OR22725.
Computer time was provided through a DOE INCITE award, number ENP008,
and an NSF XSEDE allocation award, number TG-PHY110044.
Additional computer time through an ASD allocation at NCAR is also gratefully acknowledged.
PDM is a member of the Carrera del Investigador Cient\'{\i}fico of CONICET.
Support for AP, from LASP and Bob Ergun, is gratefully acknowledged.
\end{acknowledgments}

\bibliography{ap_2014_sept_12}

\end{document}